# Electric Field Tunable Topological Phases in Graphene Nanoribbons

Fangzhou Zhao[1,2], Ting Cao[1,2,3] and Steven G. Louie[1,2]*

[1]Department of Physics, University of California at Berkeley, Berkeley, California 94720, USA

[2]Materials Sciences Division, Lawrence Berkeley National Laboratory, 1 Cyclotron Road, Berkeley, California 94720, USA.

[3]Department of Materials Science and Engineering, University of Washington, Seattle, Washington 98195, USA.

* sglouie@berkeley.edu

**Abstract:**

Graphene nanoribbons (GNRs) possess distinct symmetry-protected topological phases. We show, through first-principles calculations, that by applying an experimentally accessible transverse electric field (TEF), certain boron and nitrogen periodically co-doped GNRs have tunable topological phases. The tunability arises from a field-induced band inversion due to an opposite response of the conduction- and valance-band states to the electric field. With a spatially-varying applied field, segments of GNRs of distinct topological phases are created, resulting in a field-programmable array of topological junction states, each may be occupied with charge or spin. Our findings not only show that electric field may be used as an easy tuning knob for topological phases in quasi-one-dimensional systems, but also provide new design principles for future GNR-based quantum electronic devices through their topological characters.

The topology of a crystal's electronic structure, in par with its band structure and electron filling, plays an essential role in the properties of the system [1-3]. For example, joining two insulators of different topological classes produce robust junction states in the bandgap of these insulators [2-6]. Recently, a wide range of graphene nanoribbons (GNRs), including the armchair, cove-edged, and chevron GNRs, have been shown to host rich electronic topological phases depending on their width, edge shape and end terminations [7-10]. Moreover, the recent rapid development of bottom-up synthesis of GNRs from precursor molecules enables atomically precise design of a large variety of GNRs, including control of widths [11-13], dopant atoms [14,16], and diverse edge shapes [17-20]. Such synthesis capabilities have led to the striking experimental discovery of 1D superlattices formed by alternating segments of topologically distinct GNRs which have been measured to host one-dimensional array of topological junction states [21, 22], as predicted by theory [7].

Having the ability to controllably tune the topological invariants of materials is an actively pursued topic since it opens new opportunities for scientific studies and applications. Despite several proposals on switching between normal insulators and topological insulators (TIs) in 2D and 3D based on first-principles calculations, using external electric fields [23], tensile strains [24, 25], temperature and alloying [26-29], and so on, strategies of tuning topological phases in 1D system remain relatively underexplored. In this work, by means of first-principles calculations, we discover that topological phases of certain quasi-1D systems may be practically tuned by a new strategy that exploits external transverse electric fields (TEF). We demonstrate this strategy using a designed GNR periodically co-doped with nitrogen and boron. (See Fig. 1(a).) Topological junction states are generated by a spatially varying TEF that creates distinct topological phases in different GNR segments along the ribbon.

The band topological invariant of a 1D insulating crystal with multiple atoms per unit cell depends on the assignment of its unit cell, which in turn is dictated by the atomic structure at the end termination of the 1D system (i.e., the unit cell has to be commensurate with the boundary geometry) [7]. For instance, using the approximate chiral symmetry (the A/B sublattice symmetry) of the GNRs, the band topology of this multi-band system can be characterized by a winding number $Z$, which may be obtained using the *difference* between the intercell part of the Zak phase contributed by the A sublattice and that contributed by the B sublattice, summed over all bands up to the charge neutrality gap [10]. For such a system with a winding number $Z$, there will be $Z$ topologically protected localized in-gap state at its end termination with vacuum, according to the bulk-edge correspondence [30]. On the other hand, a $Z_2$ topological classification can be exactly applied to an 1D insulating crystal with spinless time reversal symmetry (TRS) and spatial inversion/mirror symmetry [7].

In an 1D crystal, the Zak phase [31] for the *n*th band is defined as the integral of the Berry's connection across the 1D Brillouin zone (BZ): $\gamma_n = i(\frac{2\pi}{d}) \int_{-\pi/d}^{\pi/d} dk \langle u_{nk} | \frac{\partial u_{nk}}{\partial k} \rangle$, where $u_{nk}$ is the lattice-periodic part of the Bloch state, $d$ is the unit cell size, and $k$ is the wavevector. The Zak phase of an isolated band of a general 1D insulator can take on any value, depending on the choice of the shape and origin of the unit cell. Nevertheless, if the unit cell of the crystal has inversion/mirror symmetry, the intercell (origin independent) part of the Zak phase is uniquely determined for a given unit cell shape and is quantized to 0 or $\pi$ (mod $2\pi$) [7, 8]. The $Z_2$ invariant of a 1D insulator with such symmetries is then given by $(-1)^{Z_2} = e^{i\Sigma_{n\in occ}\gamma_n}$, where the sum is over the occupied bands. When the total intercell Zak phase is $\pi$ (0) (mod $2\pi$), the $Z_2$ invariant is 1 (0). As shown in previous work [4, 7], for an unit cell with inversion $\hat{I}$ or mirror $\hat{M}$ symmetry ($\hat{O} = \hat{I}$ or $\hat{M}$), the $Z_2$ invariant of a GNR may be determined by the product of the eigenvalues of $\hat{O}$ of the states at all the time reversal invariant momentum (TRIM) k-points in the occupied band manifold: $(-1)^{Z_2} = \prod_{n\in occ} \prod_{\Gamma_i} \langle \psi_{n\Gamma_i} | \hat{O} | \psi_{n\Gamma_i} \rangle$ where the TRIM $\Gamma_i = \Gamma, X$ in the 1D BZ. In this study, we shall analyze our system using both classification schemes since it possesses TRS and spatial symmetries, as well as, to a very high degree, chiral symmetry.

For experimentally bottom-up synthesized GNR systems, the dangling $\sigma$ orbitals of the edge carbon atoms are capped by hydrogen, and are removed in energy from the bandgap region. As a result, these $\sigma$ states are not involved in the formation of the end or junction in-gap states. Also, because the GNRs considered in this work have a mirror symmetry with respect to the carbon basal plane, the $\sigma$ (mirror even) and $\pi$ (mirror odd) bands do not hybridize. Only the $\pi$ bands account for the in-gap physics of interest. Thus, both the $Z$ and $Z_2$ invariants of the GNRs of interest are calculated from the occupied band manifold of $\pi$ electrons (denoted by "$\pi$ & $occ$") only. For example, $Z_2$ is given by:

$$(-1)^{Z_2} = \prod_{n\in \pi\ \&\ occ} \langle \psi_{n\Gamma} | \hat{O} | \psi_{n\Gamma} \rangle \langle \psi_{nX} | \hat{O} | \psi_{nX} \rangle, \qquad (1)$$

In general, a transition from one topological phase to another one for an insulator requires its bandgap to close and reopen by some external tuning parameter while preserving the symmetries desired. In our case, a TEF along the width of the GNR (the y-axis) as shown in Fig. 1(a) is used because it preserves both the approximate chiral symmetry in the GNR (see

Supplementary Information (SI)) and the mirror symmetry of the GNR unit cell, allowing us to use both classification schemes.

As indicated by Eq. (1), changing the $Z_2$ invariant by band reordering at the fundamental bandgap requires firstly the wavefunctions at the minimum of the bottom conduction band (BCB) and the maximum of the top valance band (TVB) to have opposite parities at one of the $\Gamma_i$ points, i.e., $\langle\psi_{BCB\Gamma_i}|\widehat{M}|\psi_{BCB\Gamma_i}\rangle\langle\psi_{TVB\Gamma_i}|\widehat{M}|\psi_{TVB\Gamma_i}\rangle = -1$. This would ensure zero wavefunction mixing between the two states at this $\Gamma_i$ point of the BCB and TVB as a function of the TEF strength; so, a bandgap closing is ensured during the process of the induced band inversion. Secondly, to make possible a field-induced band inversion, the BCB and TVB should have opposite energy shift in response to the applied TEF, which requires the wavefunction amplitude of BCB and TVB to be asymmetric along the transverse (width-wise) direction of the GNR, i.e., the desired GNR structure should not have mirror plane that is perpendicular to the ribbon plane and the y-axis [Fig. 1(a)].

We satisfy these requirements by designing an armchair GNR (AGNR) that consists of periodic arrays of substitutional boron-dimer and nitrogen-dimer dopants, as depicted in Fig. 1(a). From our density functional theory (DFT) calculations, comparing results for an isolated boron-dimer to a nitrogen-dimer substitutionally doped onto the backbone of an AGNR shows that these two dimer defects introduce dopant states of opposite parity in the fundamental bandgap of the pristine AGNR. We therefore incorporate both boron- and nitrogen-dimer arrays into the same AGNR, and achieve having a boron-dimer (nitrogen-dimer) dopant band as its new BCB (TVB) with -1 (+1) parity eigenvalue at both $\Gamma$ and $X$, satisfying the first condition above. To satisfy the second condition, we put the boron- and nitrogen-dimer dopants near the opposite edges of the AGNR [Fig. 1(a)].

The boron- and nitrogen-dimer dopants are symmetric to a mirror plane (red dashed vertical line in Fig. 1(a)) bisecting the unit cell in the x direction which retains the mirror symmetry of the system. In addition, in AGNRs with odd number of rows of atom forming the width, the boron- and nitrogen-dimer exchange positions upon a reflection with respect to the perpendicular plane at the central backbone, defined by the blue dashed line in Fig. 1(a). We show in the SI that this ensures the system to have chiral symmetry within the nearest-neighbor tight-binding model, with and without the presence of the TEF. This allows us to classify the topology of the system by either a $Z_2$ index (using the former) or a $Z$ index (using the latter).

Among a series of GNR structures designed, guided by the above design principles, a AGNR having 11 rows of carbon atoms with one boron-dimer dopant and one nitrogen-dimer dopant in every 3 pentacene units (abbreviated as B&N-11AGNR) [Fig. 1(a)] is found to have the desired properties for field-tunable topological phases. Figure 1 depicts the Kohn-Sham band structure of this system calculated using DFT within the local density approximation (LDA) as implemented in the QUANTUM ESPRESSO (QE) package [32]. The pristine B&N-11AGNR without any applied field has a direct bandgap of ~2.9 meV. This small gap (dictated by the ribbon width [33], density and exact positions of dopants) makes an electric-field-induced band inversion experimentally feasible.

The evolution of the DFT-LDA band structure with different applied TEFs [Fig. 2] is calculated using a supercell method that has a saw-tooth potential changing along the y direction and accounts for dipole correction and depolarization field appropriately [34]. The direction from the nitrogen-dimer site to the boron-dimer site is defined as the positive direction for the TEF. We find that band inversion at the $X$ point happens at a critical field strength of the TEF of $E_c$ ~ -0.2 V/nm. For TEF with $E < E_c$, the orbital characters and parity eigenvalues of the bottom of the BCB and the top of the TVB at the $X$ point switch with each other, giving rise to inverted bands. A wavefunction projection analysis shows that, for $E < E_c$, the orbital character of the band states, as a function of wavevector k, does recover to its original character at some distance away from the $X$ point [Fig. 2(c, d)]. From Eq. (1), we obtain that the value of $Z_2$ changes from 1 to 0 as E goes below $E_c$. We also evaluate the $Z_2$ invariant by calculating the center of the Wannier functions [35-38] using the WANNIER90 package [39] (see SI). Moreover, we show that the DFT Hamiltonian for our system may be mapped approximately to a chiral Hamiltonian in a maximally localized Wannier function (MLWF) basis (see SI), and that its $Z$ index changes from 1 to 0 as the E goes below $E_c$, consistent with the above $Z_2$ classification. Thus, the B&N-11AGNR satisfies all our designing principles and has a topological $Z_2$ invariant and a $Z$ invariant that are tunable with an experimental realizable TEF in the order of 0.1 V/nm.

We next investigate the topological end states of a B&N-11AGNR finite-length segment and the relation between the number of the end states and the value of $Z$, namely, the bulk-boundary correspondence [6, 10, 30]. We perform DFT calculation of the electronic structure of a finite-length B&N-11AGNR including 24 repeating unit cells of the form shown in Fig. 1(a), with the SIESTA package [40] using a limited single zeta atomic basis. All the dangling $\sigma$ bonds at the end of the ribbon (which is a zigzag termination) are capped by hydrogen atoms. Figure

3(e) shows the evolution of bandgap and number of in-gap end states *at one end* in the charge neutrality gap as a function of the strength and direction of the TEF, which have three regions of distinct behavior. For E > $E_c$, (with $E_c$ ~ -0.8 V/nm for the critical field for band inversion from the SIESTA calculation) one topological in-gap end state appears at each end, while for $E_s$ < E < $E_c$, there is no in-gap end state ($E_s$ ~ -1.8 V/nm). For E < $E_s$, two localized states emerge in the bandgap at each end. However, these two end states, unlike the topologically protected one in the case of E > $E_c$, can be eliminated by small perturbations on the end atoms (see SI), so they are trivial end states. Thus, as expected, the bulk-boundary correspondence holds in this system. The planewave basis set used in the QE calculation of the periodic system [Fig. 2] is well converged. However, for the finite segment SIESTA calculations [Fig. 3], a limited basis set was used (because of the large number of atoms), leading to less converged bandgap values. This results in different values of $E_c$ and $E_s$ obtained by the two packages. Nevertheless, the topological character of the bands of the two calculations remains the same.

From our theory, the value of the TEF not only changes the number of end states, it also controls how localized the end states are. The local density of states (LDOS) at the end unit cell of the finite segment (integrated over the red rectangle in Fig. 3(d)) for 3 different TEFs [Fig. 3(a-c)] are compared with the bulk density of states (DOS) per unit cell. For E = 1.09 V/nm, the system has $Z = 1$, the end-cell LDOS shows a sharp and big peak at $E - E_F = 0$, arising from a very localized in-gap state at each end [Fig. 3(a)]. A smaller bulk gap at E = 0 V/nm makes the end states less localized, resulting in a lower peak height at $E - E_F = 0$ [Fig. 3(b)]. For E = -1.09 V/nm, the system has $Z = 0$. The zoom-in inset in Fig. 3(c) shows no in-gap peak in the end-cell LDOS, i.e. no in-gap end state.

A field switchable topological phase enables the use of a spatially varying TEF to create and confine topological junction states between two insulating regions of different $Z$ invariants. To illustrate this effect, we apply a superlattice electric field (i.e., a periodically repeated pattern of TEF with a repeating unit profile shown in Fig. 4(a)) on a B&N-11AGNR as shown in Fig. 4(b). The TEF is negative (zero) in the left (right) half of the supercell, and has a form $E_y = \frac{E_0}{\pi} \left[ arctan(x) - arctan\left(x - \frac{d_{sc}}{2}\right) - arctan\left(x + \frac{d_{sc}}{2}\right) - \frac{\pi}{2} \right]$, where the length of the supercell along the x direction is $d_{sc} = 305.2$ Å. A TEF of such pattern may be created by a periodic array of parallel gates with alternating bias voltages. The potential only changes rapidly between two neighboring gates in a region of ~8 Å wide. For example, a topological junction state is confined at the junction between the left part with E = -1.58 V/nm and the right part with E = 0 V/nm ($E_0$ = 1.58 V/nm) [Fig. 4(b)] because the Z invariant changes by 1 across the junction under this field profile. Because each supercell has *two* junctions, each hosting a protected junction state, the topological junction states form 2 bands with small dispersions (~5 meV) inside the common gap of the left and right "bulk" region. Hence the LDOS of the unit cell at the junction (the red rectangle in Fig. 4(b)) shows two ~5 meV wide peaks inside the common bulk gap [Fig. 4(c)]. In contrast to heterostructures of geometrically different GNRs of distinct topological phases in which junction states appear at the junction [7, 8, 10, 21, 22], the topological junction states of the B&N-11AGNR here are created by the profile of the TEF and can be moved freely to different locations in the material by varying the field profile. This gives us another degree of freedom in the rational control of topological junction states in 1D systems [7]. Furthermore, the coupling between two nearby localized states of the junction array is given by the junction separation, which is programmable by the spatial profile of the field.

In conclusion, we have proposed a scheme for designing GNRs with tunable topological phase by an applied TEF and demonstrated its feasibility through *ab initio* studies of GNRs. Our analyses and first-principles calculations show that topological end states can be created/annihilated by a uniform applied electric field on a GNR finite segment, and topological junction states can be generated in a homogeneous GNR by applying specific profile of piecewise uniform TEFs. Our study provides a new way of controlling topological junction states in 1D systems and is a promising new approach for designing future GNR-based quantum electronic devices.

Acknowledgement: We thank Z. Li, J. Jiang and M. Wu for helpful discussions. This study was supported by NSF Grant No. DMR-1926004, the NSF Center for Energy Efficient Electronics Science (E3S, NSF Grant No. ECCS-0939514), and the Office of Naval Research MURI under Award No. N00014-16-1-2921. T.C. acknowledges support from the Micron Foundation. Computational resources were provided by the DOE at Lawrence Berkeley National Laboratory's NERSC facility and the NSF through XSEDE resources at NICS.

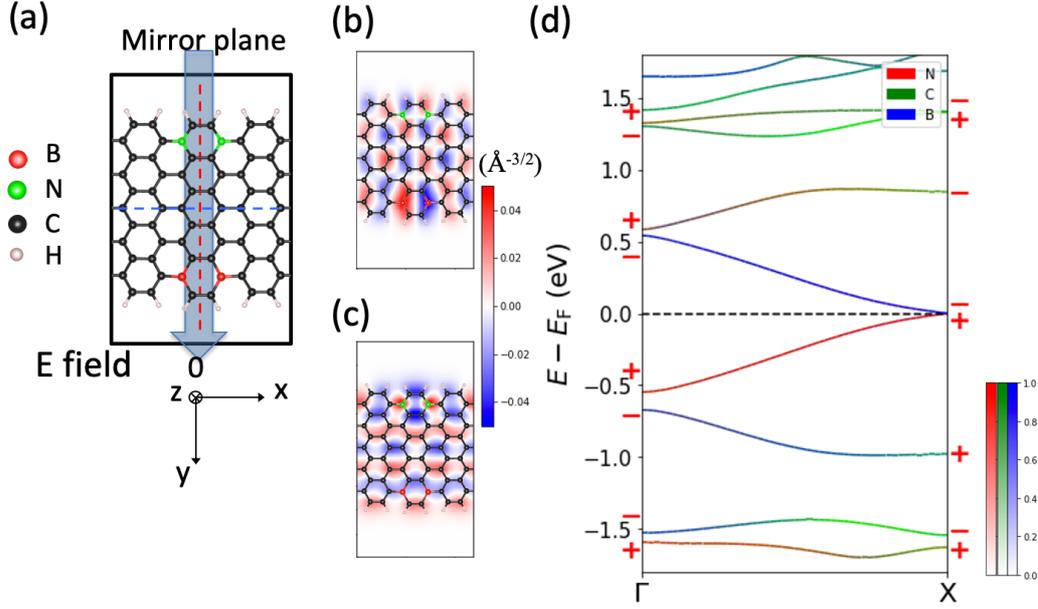

FIG. 1. The electronic structure of B&N-11AGNR. (a) A unit cell of the B&N-11AGNR that is commensurate with a zigzag end termination. The blue arrow shows the direction of a positive electric field. The red dashed line shows a mirror plane of the system, and the blue dashed line defines a normal plane about which the positions of the boron- and nitrogen-dimers are switched upon reflection. (b, c) The BCB wavefunction (b) and TVB wavefunction (c) at the Γ point, plotted on a plane at 1 Å above the GNR basal plane. (d) The band structure of B&N-11AGNR without any applied field. The blue, red and green colors in the band structure denote the module squared weights of the wavefunction that are projected onto the boron, nitrogen and carbon atomic orbitals, respectively. The scale bar defines the mapping between the color scale and the percentage weight. The projection has been normalized according to the total number of atoms of each species per unit cell. The parity eigenvalues $\langle\psi_{n\Gamma_i}|\hat{M}|\psi_{n\Gamma_i}\rangle$ of the 8 bands near the Fermi level $E_F$ at Γ and $X$ are marked as "+" and "-" for value of +1 and -1, respectively.

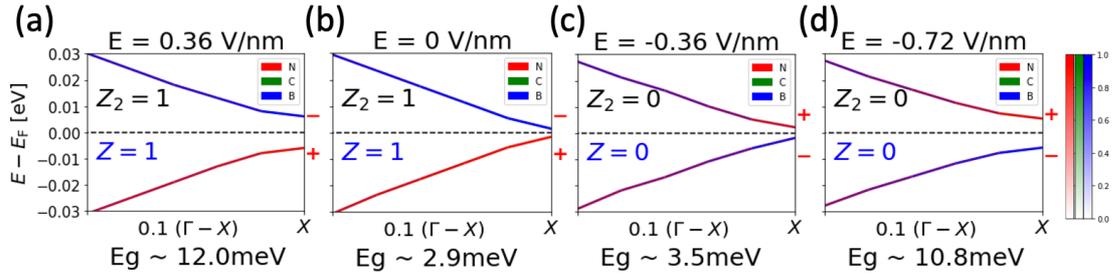

FIG. 2. The band inversion process under changing TEFs from DFT-LDA calculations. As the TEF goes below a critical field strength of $E_c \sim -0.2$ V/nm, the characters (and parities) of the states of the BCB and TVB at the $X$ point are inverted, and both the $Z$ and $Z_2$ invariants of this system change from 1 to 0. The bands are zoomed in a region near the $X$ point in reciprocal space spanning over 1/10 of the $\Gamma - X$ length.

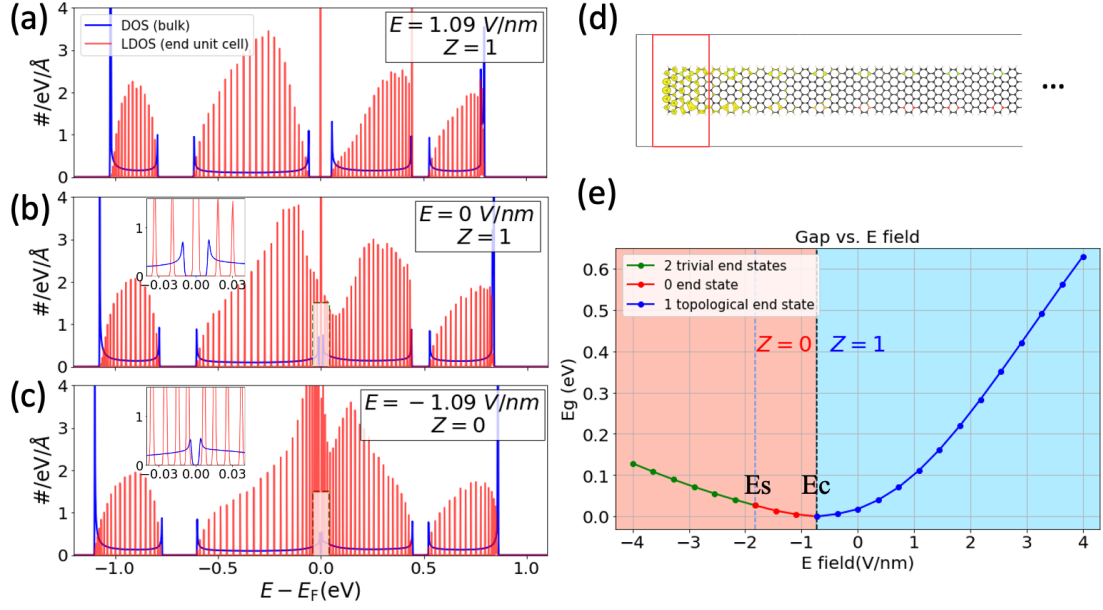

FIG. 3. Bulk-boundary correspondence. (a-c) The red curves show the calculated LDOS integrated in the end unit cell (the red rectangular region in (d)) of the 24-unit cell finite-length GNR with TEF of E = 1.09 (a), 0 (b), and -1.09 (c) V/nm, while the blue curves show the DOS per unit cell of the bulk periodic system at the same TEFs. Both the LDOS and DOS are in units of number of states per energy per length without considering spin degeneracy. The Gaussian broadening factor employed for both the LDOS and bulk DOS is 1 meV. The insets show zoom-in plots in the green dashed rectangular regions. (d) The iso-surface charge density plot at $1.4 \times 10^{-5}$ Å$^{-3}$ (1% of the maximum value) of the topological end state in the 24-unit-cell finite segment calculation (only the left 1/3 of the segment is shown). The TEF is 1.09 V/nm ($Z = 1$). (e) DFT-LDA bandgap versus TEF calculated using the SIESTA package with a limited single zeta basis. $E_c$ is at ~ -0.8 V/nm. The calculated $Z$ invariant is 1 (0) for $E > E_c$ ($E < E_c$). The colors on different parts of the curve denote the number of end states per end for the system.

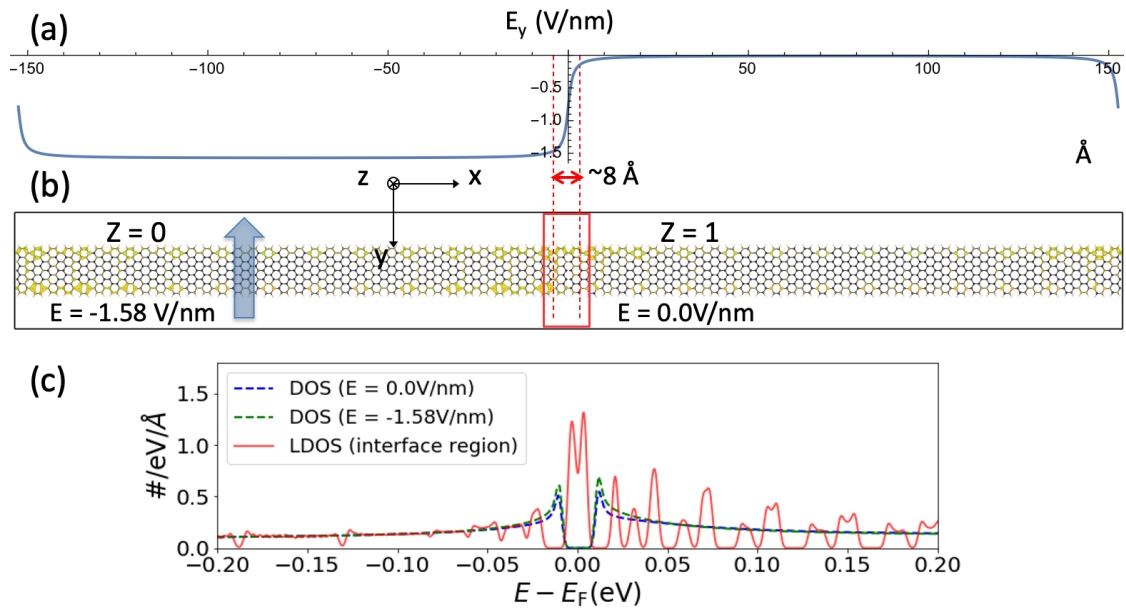

FIG. 4. (a) The TEF profile in one repeating period as a function of coordinate $x$ which is along the axis of the ribbon. (b) The iso-surface charge density plot at $2.7 \times 10^{-6} \text{Å}^{-3}$ (2% of the maximum value) of the topological junction state (evaluated at Γ-point of the superlattice) of a B&N-11AGNR with the superlattice electric field in (a) applied. One repeating period of the TEF (the supercell shown in (b)) contains 24 B&N-11AGNR unit cells. The field strengths in the center of the left (right) region are -1.58 V/nm (0 V/nm) resulting in the system with $Z$ being 0 (1). (c) DOS of the GNR in a superlattice electric field. The red solid curve shows the LDOS computed in the junction region (the red rectangle in (b)). The green (blue) dashed curves show the DOS per unit cell in bulk calculations with a uniform field TEF of E = -1.58 V/nm (0 V/nm).